\documentclass{llncs}
\usepackage{makeidx}  
\begin{document}
%
%
\pagestyle{headings}  
%
% ...
\mainmatter              
\title{Declarative Diagnosis of Floundering}
\titlerunning{Declarative Diagnosis of Floundering}  
%
\author{Lee Naish}
\authorrunning{L. Naish}   
%
\tocauthor{Lee Naish}
\institute{University of Melbourne, Melbourne 3010, Australia\\
\email{lee@csse.unimelb.edu.au},\\
\texttt{http://www.csse.unimelb.edu.au/\homedir lee/} }

\maketitle              

\begin{abstract}

Many logic programming languages have delay primitives which allow
coroutining. This introduces a class of bug symptoms --- computations
can \emph{flounder} when they are intended to succeed or finitely fail.
For concurrent logic programs this is normally called \emph{deadlock}.
Similarly, constraint logic programs can fail to invoke certain constraint
solvers because variables are insufficiently instantiated or constrained.
Diagnosing such faults has received relatively little attention to date.
Since delay primitives affect the procedural but not the declarative view
of programs, it may be expected that debugging would have to consider the
often complex details of interleaved execution.  However, recent work on
semantics has suggested an alternative approach.  In this paper we show
how the declarative debugging paradigm can be used to diagnose unexpected
floundering, insulating the user from the complexities of the execution.

\end{abstract}

\section{Introduction}
\label{sec-intro}

The first Prolog systems used a strict left to right evaluation strategy,
or computation rule.  However, since the first few years of logic
programming there have been systems which support coroutining between
different sub-goals \cite{clark79}.  Although the default order is
normally left to right, individual calls can \emph{delay} if certain
arguments are insufficiently instantiated, and later \emph{resume}, after
other parts of the computation have further instantiated them.  Such
facilities are now widely supported in Prolog systems.  They also gave
rise to the class of \emph{concurrent} logic programming languages, such
as Parlog \cite{Gre87}, where the default evaluation strategy is parallel
execution and similar delay mechanisms are used for synchronisation and
prevention of unwanted nondeterminism.  Delay mechanisms have also been
influential for the development of \emph{constraint logic programming}
\cite{JafLas87}.  Delays are often used when constraints are ``too hard''
to be handled by efficient constraint solvers, for example, non-linear
constraints over real numbers.


Of course, more features means more classes of bugs.  In theory delays
don't affect soundness of Prolog (see \cite{Llo84}\footnote{In practice,
floundering within negation can cause unsoundness.}) --- they can be seen
as affecting the ``control'' of the program without affecting the logic
\cite{kowalski}.  However, they do introduce a new class of bug symptoms.
A call can delay and never be resumed (because it is never sufficiently
instantiated); the computation is said to \emph{flounder}.  Most Prolog
systems with delays still print variable bindings for floundered
derivations in the same way as successful derivations (in this paper
we refer to these as ``floundered answers''), and may also print some
indication that the computation floundered.  Floundered answers are not
necessarily valid, or even satisfiable, according to the declarative
reading of the program.  They provide little useful information and
generally indicate the presence of a bug.  In concurrent logic programs
the equivalent of floundering is normally called \emph{deadlock} ---
the computation terminates with no ``process'' (call) sufficiently
instantiated to proceed.  In constraint logic programming systems,
the analogue is a computation which terminates with some insufficiently
instantiated constraints not solved (or even checked for satisfiability).
Alternatively, if some constraints are insufficiently instantiated they
may end up being solved by less efficient means than expected, such as
exhaustive search over all possible instances.

There is a clear need for tools and techniques to help diagnose
floundering in Prolog (and analogous bug symptoms in other logic
programming languages), yet there has been very little research in
this area to date.  There has been some work on showing floundering
is impossible using syntactic restrictions on goals and programs
(particularly logic databases), or static analysis methods (for
example, \cite{MSD90}\cite{delay-popl}).  However, this is a far cry
from general purpose methods for diagnosing floundering.  In this paper
we present such a method.  Furthermore, it is a surprisingly attractive
method, being based on the declarative debugging paradigm \cite{Sha83}
which is able to hide many of the procedural details of a computation.
The paper is structured as follows.  We first give some examples of how
various classes of bugs can lead to floundering.  We then present our
method of diagnosing floundering, give examples, and discuss how our
simple prototype could be improved.  Next we briefly consider some more
theoretical aspects, then conclude.  Basic familiarity of Prolog with
delays and declarative debugging is assumed.

\section{Example}
\label{sec-example}

\begin{figure}
\begin{center}
\begin{verbatim}
% perm(As0, As): As = permutation of list As0
% As0 or As should be input
perm([], []).
perm([A0|As0], [A|As]) :-
    when((nonvar(As1) ; nonvar(As)),
        inserted(A0, As1, [A|As])),
    when((nonvar(As0) ; nonvar(As1)),
        perm(As0, As1)).

% inserted(A, As0, As): As = list As0 with element A inserted
%......% As0 should be input         %... Bug 2
% As0 or As should be input
inserted(A, As0, [A|As0]).
inserted(A, [A1|As0], [A1|As]) :-
%...when(nonvar(As0),                %... Bug 2
%...when((nonvar(As0) ; nonvar(A)),  %... Bug 1
    when((nonvar(As0) ; nonvar(As)),
%.......inserted(A, AS0, As)).       %... Bug 3
        inserted(A, As0, As)).
\end{verbatim}
\end{center}
\caption{A reversible permutation program}
\label{fig_perm}
\end{figure}


Figure \thefigure\ gives a permutation program which has simple logic but
is made reversible by use of delaying primitives and careful ordering of
sub-goals in \texttt{perm/2} (see \cite{naish:thesis:86} for further
discussion).  The delay primitive used is the ``when meta-call'':
a call \texttt{when(Cond,A)} delays until condition \texttt{Cond}
is satisfied, then calls \texttt{A}.  For example, the recursive call
to \texttt{perm/2} will delay until at least one of its arguments are
non-variables.  Generally there are other features supported, such as
delaying until a variable is ground; we don't discuss them here, though
our method and prototype support them.  A great number of delay primitives
have been proposed.  Some, like the when meta-call, are based on calls.
Others are based on procedures (affecting all calls to the procedure),
which is often more convenient and tends to clutter the source code less.
Our general approach to diagnosis is not effected by the style of delay
primitive.  The when meta-call is by far the most portable of the more
flexible delay primitives, which is our main reason for choosing it.
We have developed the code in this paper using SWI-Prolog.

We consider three separate possible bugs which could have been
introduced, shown as commented-out lines preceding the correct versions.
They exemplify three classes of errors which can lead to floundering:
incorrect delay annotations, confusion over the modes of predicates,
and logical errors.  With the first bug, an incorrect delay annotation
on the recursive call to \texttt{inserted/3}, several bug symptoms
are exhibited.  The call \verb@perm([X,Y,Z],A)@ behaves correctly but
\verb@perm([1,2,3],A)@ succeeds with the answers \verb@A=[1,2,3]@
and \verb@A=[1,3,2]@, then loops indefinitely.  We don't consider
diagnosis of loops in this paper, though they are an important symptom
of incorrect control.  The call \verb@perm(A,[1,2,3])@ succeeds with
the answer \verb@A=[1,2,3]@ then has three further floundered answers,
\verb@A=[1,2,_,_|_]@, \verb@A=[1,_,_|_]@ and \verb@A=[_,_|_]@, before
terminating with failure.

The second bug is a more subtle control error.  When \texttt{inserted/3}
was coded we assume the intention was the second argument should always be
input and the delay annotation is correct with respect to this intention.
However, some modes of \texttt{perm/2} require \texttt{inserted/3} to
work with just the third argument input.  When coding \texttt{perm/2}
the programmer was either unaware of this or was confused about what
modes \texttt{inserted/3} supported.  Although this version of the
program behaves identically to Bug 1 for the goal \verb@perm(A,[1,2,3])@,
the bug diagnosis will be different because the programmer intentions
are different.  The mistake was made in the coding of \texttt{perm/2},
and this is reflected in the diagnosis.  The simplest way to \emph{fix}
the bug is change the intentions and code for \texttt{inserted/3},
but we only deal with diagnosis in this paper.

The third bug is a logical error in the recursive call to
\texttt{inserted/3}.  Due to an incorrect variable name, other variables
remain uninstantiated and this can ultimately result in floundering.  The
call \verb@perm([1,2,3],A)@ first succeeds with answer \verb@A=[1,2,3]@.
There are four other successful answers which are satisfiable but not
valid, for example, \verb@A=[1,2,3|_]@ and \verb@A=[3,1|_]@.  These could
be diagnosed by existing wrong answer declarative debugging algorithms,
though some early approaches assumed bug symptoms were unsatisfiable
atoms (see \cite{ddscheme}).  These answers are interleaved with
four floundered answers, such as \verb@A=[1,3,_|_]@, which are also
satisfiable but not valid.  The call \verb@perm(A,[1,2,3])@ succeeds
with the answer \verb@A=[1,2,3]@ then has three floundered answers, also
including \verb@A=[1,3,_|_]@.  The call \verb@perm([A,1|B],[2,3])@ should
finitely fail but returns a single floundered answer with \verb@A=3@.

Because delays are the basic cause of floundering and they are inherently
procedural, it is natural to assume that diagnosing unexpected floundering
requires a procedural view of the execution.  Even with such a simple
program and goals, diagnosis using just traces of floundered executions
can be extremely difficult.  Subcomputations may delay and be resumed
multiple times as variables incrementally become further instantiated.
Reconstructing how a single subcomputation proceeds can be very difficult,
especially if there is also backtracking involved.  Although some tools
have been developed, such as printing the history of instantiation states
for a variable, diagnosis of floundering has remained very challenging.

\section{Declarative diagnosis of floundering}
\label{sec-dd}

To diagnose unexpected floundering in pure Prolog programs with delays
we use an instance of the three-valued declarative debugging scheme
described in \cite{ddscheme3}. We describe the instance precisely
in the following sections, but first introduce the general scheme. A
computation is represented as a tree, with each node associated with
a section of source code (a clause in this instance) and subtrees
representing subcomputations.  The trees we use here are a generalisation
of proof trees.  Each node has a truth value which expresses how
the subcomputation compares with the intentions of the programmer.
Normally the truth values of only some nodes are required and are found by
asking the user questions.  Three truth values are used: \emph{correct},
\emph{erroneous}, and \emph{inadmissible}.  Informally, the third truth
value means the subcomputation should never have occurred.  It means a
\emph{pre-condition} of the code has been violated, whereas erroneous
means a \emph{post-condition} has been violated.  Inadmissibility was
initially used to express the fact that a call was ill-typed \cite{Per86}
but can also be used for other purposes \cite{ddscheme3}.  Here calls
which flounder because they never become sufficiently instantiated are
considered inadmissible.

Given a tree with truth values for each node, a node is \emph{buggy}
if it is erroneous but has no erroneous children.  Diagnosis consists of
searching the tree for a buggy node.  Many search strategies are possible
and \cite{ddscheme3} provides very simple code for a top-down search.
The code first checks that the root is erroneous.  It then recursively
searches for bugs in children and returns them if they exist.  Otherwise
the root is returned as a buggy node, along with an inadmissible child
if any are found.  In the next sections we first define the trees we
use, discuss how programmer intentions are formalised, give some simple
diagnosis sessions then make some remarks about search strategy.

\subsection{Partial proof trees}

Standard wrong answer declarative diagnosis uses Prolog proof trees
which correspond to successful derivations (see \cite{Llo84}).  Each node
contains an atomic goal which was proved in the derivation (in its final
state of evaluation) and the children of a node are the subgoals of the
clause used to prove the goal.  Leaves are atomic goals which were matched
with unit clauses.  We use \emph{partial} proof trees which correspond
to successful \emph{or floundered} derivations.  The only difference is
they have an additional class of leaves: atomic goals which were never
matched with any clause because they were delayed and never resumed.

\begin{definition}[(Callable) annotated atom]
An \emph{annotated atom} is an atomic formula or a term of the form
$when(C,A)$, where $A$ is an atomic formula and $C$ is a condition of a
when meta-call.
It is \emph{callable} if it is an atom or
$C$ is true according to the normal Prolog
meaning (for ``,'', ``;'' and \texttt{nonvar/1}).
$atom(X)$ is the atom of annotated atom $X$.
\end{definition}

\begin{definition}[(Successful or floundered) partial proof tree]
A \emph{partial proof tree} for annotated atom $A$ and program $P$ is either
\begin{enumerate}
\item a node containing $A$, where $atom(A)$ is an instance of
a unit clause in $P$ or $A$ is not callable, or
\item a node containing $A$ together with partial proof
(sub)trees $S_i$ for annotated atom $B_i$ and $P$, $i = 1 \ldots n$,
where $atom(A) \texttt{:-} B_1, \ldots B_n$ is an instance of a clause in $P$.
\end{enumerate}
A partial proof tree is \emph{floundered} if it contains any
annotated atoms which are not callable, otherwise it is
\emph{successful}.
\end{definition}

\begin{figure}
\begin{center}
\begin{verbatim}
% solve_atom(A, C0, C, AT): A is an atomic goal, possibly wrapped
% in when meta-call, which has succeeded or floundered;
% AT is the corresponding partial proof tree with floundered
% leaves having a variable as the list of children;
% C0==C if A succeeded
solve_atom(when(Cond, A), C0, C, AT) :- !,
    AT = node(when(Cond, A), C0, C, Ts),
    when(Cond, solve_atom(A, C0, C,  node(_, _, _, Ts))).
solve_atom(A, C0, C, node(A, C0, C, AsTs)) :-
    clause(A, As),
    solve_conj(As, C0, C, AsTs).

% As above for conjunction; returns list of trees
solve_conj(true, C, C, []) :- !.
solve_conj((A, As), C0, C, [AT|AsTs]) :- !,
    solve_atom(A, C0, C1, AT),
    solve_conj(As, C1, C, AsTs).
solve_conj(A, C0, C, [AT]) :-
    solve_atom(A, C0, C, AT).
\end{verbatim}
\end{center}
\caption{A meta-interpreter which builds partial proof trees}
\label{fig_solve}
\end{figure}

Declarative debuggers use various methods for representing trees and
building such representations.  The declarative debugger for Mercury
\cite{merc_debug} is a relatively mature implementation. A much simpler
method (which is impractical for large scale applications) is a meta
interpreter which constructs an explicit representation of the tree.
Figure \thefigure\ is one such (poor) implementation which we include
for completeness.  Floundering is detected using the ``short circuit''
technique --- an accumulator pair is associated with each subgoal and the
two arguments are unified if and when the subgoal succeeds.  Tree nodes
contain an annotated atom, this accumulator pair and a list of subtrees.
A subcomputation is floundered if the accumulator arguments in the root
of the subtree are not identical.

\subsection{The programmer's intentions}

The way truth values are assigned to nodes encodes the user's intended
behaviour of the program.   For traditional declarative debugging of
wrong answers the intended behaviour can be specified by partitioning the
set of ground atoms into true atoms and false atoms.  There can still be
non-ground atoms in proof tree nodes, which are considered true if the
atom is \emph{valid} (all instances are true).  A difficulty with this
two-valued scheme is that most programmers make implicit assumptions
about they way their code will be called, such as the ``type''
of arguments.  For example, it is assumed that \texttt{inserted/3}
will be called in a context where (at least one of) the last two arguments
must be lists.  Although \verb@inserted(1,a,[1|a])@ can succeed, it is
counter-intuitive to consider it to be true (since it is ``ill-typed''),
and if it is considered false then the definition of \texttt{inserted/3}
must be regarded as having a logical error.  The solution to this
problem is to be more explicit about how predicates should be called,
allowing pre-conditions \cite{drabent} or saying that certain things are
inadmissible \cite{Per86} or having a three-way partitioning of the set
of ground atoms \cite{naish:sem3neg}.

In the case of floundering the intended behaviour of non-ground atoms
must be considered explicitly.  As well as assumptions about types
of arguments, we inevitably make assumptions about how instantiated
arguments are.  For example, \texttt{perm/2} is not designed to generate
all solutions to calls where neither argument is a (nil-terminated)
list and even if it was, such usage would most likely cause an infinite
loop if used as part of a larger computation.  It is reasonable to say
that such a call to \texttt{perm/2} should not occur, and hence should
be considered inadmissible, even though more instantiated calls are
acceptable.  An important heuristic for generating control information is
that calls which have an infinite number of solutions should be avoided
\cite{naish:thesis:86}.  Instead, such a call is better delayed, in the
hope that other parts of the computation will further instantiate it and
make the number of solutions finite.  If the number of solutions remains
infinite the result is floundering, but this is still preferable to an
infinite loop.

We specify the intended behaviour of a program as follows:
\begin{definition}[Interpretation]
An \emph{interpretation} is a three-way partitioning of the set
of all atoms into those which are \emph{inadmissible}, \emph{valid} and
\emph{erroneous}.  The set of admissible (valid or erroneous) atoms is
closed under instantiation (if an atom is admissible then any instance
of it is admissible), as is the set of valid atoms.
\end{definition}
In our example \texttt{perm(As0,As)} is admissible if and only if either
\texttt{As0} or \texttt{As} are (nil-terminated) lists, and valid if and
only if \texttt{As} is a permutation of \texttt{As0}.  This expresses the
fact that either of the arguments can be input, and only the list skeleton
(not the elements) is required.  For example, \verb@perm([X],[X])@ is
valid (as are all its instances), \verb@perm([X],[2|Y])@ is admissible
(as are all its instances) but erroneous (though an instance is valid) and
\verb@perm([2|X],[2|Y])@ is inadmissible (as are all atoms with this as
an instance).  For diagnosing Bug 2, we assume \texttt{inserted(A,As0,As)}
is admissible if and only if \texttt{As0} is a list.  For diagnosing the
other bugs either \texttt{As0} or \texttt{As} are lists, expressing the
different intended modes in these cases.


Note we do not have different admissibility criteria for different
sub-goals in the program --- the intended semantics is predicate-based.
Delay primitive based on predicates thus have an advantage of being
natural from this perspective.  Note also that atoms in partial
proof tree nodes are in their final state of instantiation in the
computation.  It may be that in the first call to \texttt{inserted/3}
from \texttt{perm/2}, no argument is instantiated to a list (it may delay
initially), but as long as it is eventually sufficiently instantiated
(due to the execution of the recursive \texttt{perm/2} call, for example)
it is considered admissible.  However, since admissibility is closed
under instantiation, an atom which is inadmissible in a partial proof
tree could not have been admissible at any stage of the computation.
The debugger only deals with whether a call flounders --- the lower level
procedural details of when it is called, delayed, resumed \emph{et cetera}
are hidden.

Truth values of partial proof tree nodes are defined in terms of the
user's intentions:
\begin{definition}[Truth of nodes]
Given an interpretation $I$, a partial proof tree node is
\begin{enumerate}
\item \emph{correct}, if the atom in the node is valid in $I$ and
the subtree is successful,
\item \emph{inadmissible}, if the atom in the node is
inadmissible in $I$, and
\item \emph{erroneous}, otherwise.
\end{enumerate}
\end{definition}
Note that floundered subcomputations are never correct.  If the atom is
insufficiently instantiated (or ``ill-typed'') they are inadmissible,
otherwise they are erroneous.

\begin{figure}
\begin{center}
\begin{verbatim}
?- wrong(perm(A,[1,2,3])).
(succeeded)  perm([1, 2, 3], [1, 2, 3]) ...? v
(floundered) perm([1, 2, A, B|C], [1, 2, 3]) ...? e
(floundered) perm([2, A, B|C], [2, 3]) ...? e
(floundered) perm([A, B|C], [3]) ...? e
(floundered) inserted(A, [3|B], [3]) ...? e
(floundered) inserted(A, B, []) ...? e
BUG - incorrect delay annotation:
when((nonvar(A);nonvar(B)), inserted(B, A, []))
\end{verbatim}
\end{center}
\caption{Diagnosis of bug 1}
\label{fig_d1}
\end{figure}

\subsection{Diagnosis examples}

In our examples we use a top-down search for a buggy node, which
gives a relatively clear picture of the partial proof tree.  They are
copied from actual runs of our prototype\footnote{Available from
\texttt{http://www.cs.mu.oz.au/\char'176lee/papers/ddf/}} except that
repeated identical questions are removed.  In section \ref{sec-search}
we discuss strategies which can reduce the number of questions; the way
diagnoses are printed could also be improved. Figure \ref{fig_d1} shows
how Bug 1 is diagnosed.  We use a top-level predicate \texttt{wrong/1}
which takes an atomic goal, builds a partial proof tree for an instance of
the goal then searches the tree.  The truth value of nodes is determined
from the user.  The debugger prints whether the node succeeded or
floundered (this can be helpful to the user, and the reader, though it
is not necessary), then the atom in the node is printed and the user is
expected to say if it is valid (\texttt{v}), inadmissible (\texttt{i})
or erroneous (\texttt{e})\footnote{To help with missing answer diagnosis
it would be preferable to distinguish unsatisfiable atoms from those
which are satisfiable but not valid.}.  The first question relates to
the first answer returned by the goal.  It is valid, so the diagnosis
code fails and the computation backtracks, building a new partial proof
tree for the next answer, which is floundered.  The root of this tree is
determined to be erroneous and after a few more questions a buggy node
is found.  It is a floundered leaf node so the appropriate diagnosis is
an incorrect delay annotation, which causes \verb@inserted(A,B,[])@ to
delay indefinitely (rather than fail).  Ideally we should also display
the instance of the clause which contained the call (the debugger code
in \cite{ddscheme3} could be modified to return the buggy node \emph{and}
its parent), and the source code location.

\begin{figure}
\begin{center}
\begin{verbatim}
?- wrong(perm(A,[1,2,3])).
(succeeded)  perm([1, 2, 3], [1, 2, 3]) ...? v
(floundered) perm([1, 2, A, B|C], [1, 2, 3]) ...? e
(floundered) perm([2, A, B|C], [2, 3]) ...? e
(floundered) perm([A, B|C], [3]) ...? e
(floundered) inserted(A, [3|B], [3]) ...? i
(floundered) perm([A|B], [3|C]) ...? i
BUG - incorrect modes/types in clause instance:
perm([A, C|D], [3]) :-
        when((nonvar([3|B]);nonvar([])), inserted(A, [3|B], [3])),
        when((nonvar([C|D]);nonvar([3|B])), perm([C|D], [3|B])).
\end{verbatim}
\end{center}
\caption{Diagnosis of bug 2}
\label{fig_d2}
\end{figure}

Figure \ref{fig_d2} shows how Bug 2 is diagnosed.  It proceeds in a
similar way to the previous example, but due to the different programmer
intentions (the mode for \texttt{inserted/3}) the floundering call
\verb@inserted(A,[3|B],[3])@ is considered inadmissible rather than
erroneous, eventually leading to a different diagnosis.  Both calls in the
buggy clause instance are inadmissible.  The debugger of \cite{ddscheme3}
returns both these inadmissible calls as separate diagnoses.  For
diagnosing floundering it is preferable to return a single diagnosis,
since the floundering of one can result in the floundering of another
and its not clear which are the actual culprit(s).

\begin{figure}
% ...
\begin{center}
\begin{verbatim}
?- wrong(perm(A,[1,2,3])).
(succeeded)  perm([1, 2, 3], [1, 2, 3]) ...? v
(floundered) perm([1, 3, A|B], [1, 2, 3]) ...? e
(floundered) perm([3, A|B], [2, 3]) ...? e
(floundered) perm([A|B], [2|C]) ...? i
(succeeded)  inserted(3, [2|A], [2, 3]) ...? e
(succeeded)  inserted(3, [], [3]) ...? v
BUG - incorrect clause instance:
inserted(3, [2|A], [2, 3]) :-
        when((nonvar(A);nonvar([3])), inserted(3, [], [3])).
\end{verbatim}
\end{center}
\caption{Diagnosis of bug 3}
\label{fig_d3a}
\end{figure}

\begin{figure}
\begin{center}
\begin{verbatim}
...
(floundered) perm([1, 2, 3], [1, 3, A|B]) ...? e
(floundered) perm([2, 3], [3, A|B]) ...? e
(floundered) inserted(2, [3], [3, A|B]) ...? e
(floundered) inserted(2, [A|B], [A|C]) ...? i
BUG - incorrect modes/types in clause instance:
inserted(2, [3], [3, A|B]) :-
        when((nonvar([]);nonvar([A|B])), inserted(2, [A|_], [A|B])).
\end{verbatim}
\end{center}
\caption{Diagnosis of bug 3 using goal \texttt{perm([1,2,3],A)} }
\label{fig_d3b}
\end{figure}

Figures \ref{fig_d3a}\ and \ref{fig_d3b}\ show how Bug 3 is
diagnosed.  In the first case the diagnosis is a logical error in the
\texttt{inserted/3} clause.  In the second case the top-level goal
is \verb@perm([1,2,3],A)@.  We assume the user decides to diagnose a
floundered answer, skipping over the previous answers.  The diagnosis is a
control error, similar to that for Bug 2.  Both are legitimate diagnoses,
just as logical bugs can lead to both missing and wrong answers, which
typically result in different diagnoses in declarative debuggers.

\subsection{Search strategy}
\label{sec-search}

\begin{figure}
\begin{center}
\begin{verbatim}
% returns children of a node, floundered ones first
child(node(_, _, _, Ts), T) :-
    nonvar(Ts), % not a floundered leaf
    (    member(T, Ts),
         T = node(_, C0, C, _),
         C0 \== C   % T is floundered
     ;
         member(T, Ts),
         T = node(_, C0, C, _),
         C0 == C    % T is not floundered
    ).
\end{verbatim}
\end{center}
\caption{Finding children of a partial proof tree node}
\label{fig_child}
\end{figure}

We have used a very simple search strategy in our examples.  Suggestions
for search strategies for diagnosing some forms of abnormal termination
are given in \cite{ddscheme3} and these can be adapted to floundering.
From our definition of truth values for nodes, we know no floundered node
is correct.  We also know that floundering is caused by (at least one)
floundered leaf node.  Thus we have (at least one) path of nodes which
are not correct between the root node and a leaf.  It makes sense to
initially restrict our search to such a path.  A top-down search of
the path can be achieved simply by careful ordering of the children
(examining floundered children first) in a top-down debugger.  This is
what we have used for our examples (see Figure \thefigure\ for the code).
There is an erroneous node on the path with no erroneous children on
the path.  Both bottom-up and binary search strategies are likely to find
this node significantly more quickly than a top-down search. Once this
node is found, its other children must also be checked.  If there are no
erroneous children the node is buggy.  Otherwise, an erroneous child can
be diagnosed recursively, if it is floundered, or by established wrong
answer diagnosis algorithms.

\section{Theoretical considerations}
\label{sec-theory}

We first make some remarks about the soundness and completeness
of this method of diagnosis, then discuss related theoretical work.
An admissible atomic formula which flounders has a finite partial proof
tree with an erroneous root and clearly this must have a buggy node.
Since the search space is finite, completeness can easily be achieved.
Soundness criteria come from the definition of buggy nodes (erroneous
nodes with no erroneous children).  The three classes of bugs mentioned
in Section \ref{sec-example} give a complete categorisation of bugs
which cause floundering.  Logical errors cause successful buggy nodes.
Incorrect delay annotations cause floundered leaf nodes which are
admissible but delay.  Confusion over the modes causes floundered
internal nodes which are admissible but have one or more floundered
inadmissible children.  If there are also successful inadmissible
(``ill-typed'') children it may be more natural to say it is caused by
a logical (``type'') error.

Declarative diagnosis of wrong answers can hide the complex procedural
details of execution because success is independent of the computation
rule.  Our current work on diagnosis arose out of more theoretical work
on floundering \cite{flounder}.  Nearly all delay primitives have the
property that if a certain call can proceed (rather than delay), any more
instantiated version of the call can also proceed.  An important result
which follows from this property is similar to the result concerning
success: whether a computation flounders, and the final instantiation
of variables, depends on the delay annotations but not on the order in
which sufficiently instantiated call are selected.  Non-floundering is
also closed under instantiation, so it is natural for admissibility to
inherit this restriction and partial proof trees provide a basis for
intuitive diagnoses.  Our diagnosis method can be effectively applied to
other delay primitives for which this property holds simply by changing
the definition of callable annotated atoms.

The use of the term ``declarative'' in this paper may have caused
unease in some readers.  However, there is an interpretation of when
meta-calls which allows model-theoretic view of our diagnosis method (see
\cite{flounder} for further details).  We partition the set of function
symbols into \emph{program} function symbols and \emph{extraneous}
function symbols.  The program, goals and set of admissible atoms only
contain program function symbols.  We interpret \texttt{nonvar(X)}
as meaning the principle function symbol of \texttt{X} is a program
function symbol.  Instead of a when meta call \texttt{when(C,G)}
being interpreted as \texttt{G}, we interpret it as a disjunction
\texttt{(G;$\bar{\texttt{C}}$)}, where $\bar{\texttt{C}}$ is the negation
of \texttt{C}. For example, the meaning of \texttt{when(nonvar(X),p(X))}
is \texttt{p(X)} \emph{or} the principle function symbol of \texttt{X}
is extraneous.  Extraneous function symbols are essentially used to
\emph{encode} variables.

A goal has a floundered derivation which uses the normal procedural
interpretation of when meta-calls if and only if it has a successful
derivation using an added disjunct ($\bar{\texttt{C}}$) in the
alternative interpretation.  The sets of admissible and valid atoms
can also be encoded in the same way: if an atom containing variables
is admissible (or valid), the atom with the variables instantiated
to extraneous function symbols should be admissible (or valid,
respectively).  Encoding our previous example, \verb@perm([$],[$])@
would be valid, \verb@perm([$],[2|$$])@ would be erroneous and
\verb@perm([2|$],[2|$$])@ would be inadmissible, assuming \texttt{\$}
and \texttt{\$\$} are extraneous function symbols.  We then have a
partitioning of ground atoms into those which are true (valid), false, and
inadmissible --- a three-valued interpretation of the kind used discussed
\cite{naish:sem3neg}.  If this interpretation is not a three-valued model,
bug symptoms can be diagnosed using declarative wrong answer diagnosis.
All the diagnosis examples in this paper can be reproduced in this way,
though floundering of valid atoms (which is rare in practice) cannot
be diagnosed.  In this paper the way truth values are assigned to tree
nodes overcomes this limitation.

\section{Conclusion}
\label{sec-conc}

There has long been a need for tools and techniques to diagnose unexpected
floundering in Prolog with delay primitives, and related classes of
bug symptoms in other logic programming languages.  The philosophy
behind delay primitives in logic programming languages is largely based
on Kowalski's equation: Algorithm = Logic + Control \cite{kowalski}.
By using more complex control, the logic can be simpler.  This allows
simpler reasoning about correctness of answers from successful derivations
--- we can use a purely declarative view, ignoring the control because
it only affects the procedural semantics.  When there are bugs related
to control it is not clear the trade-off is such a good one. The control
and logic can no longer be separated.  Since the normal declarative view
cannot be used, the only obvious option is to use the procedural view.
Unfortunately, even simple programs can exhibit very complex procedural
behaviour, making it very difficult to diagnose and correct bugs using
this view of the program.

In the case of floundering, a much simpler high level approach turns
out to be possible.  The combination of the logic and control can be
viewed as just slightly different logic, allowing declarative diagnosis
techniques to be used.  The procedural details of calls delaying
and the interleaving of subcomputations can be ignored.  The user
can simply put each atomic formula into one of three categories. The
first is inadmissible: atoms which should not be called because they are
insufficiently instantiated and expected to flounder (or are ``ill-typed''
or violate some pre-condition of the procedure).  The second is valid:
atoms for which all instances are true and are expected to succeed.
The third is erroneous: atoms which are legitimate to call but which
should not succeed without being further instantiated (they are not valid,
though an instance may be).  A floundered derivation can be viewed as a
tree and this three-valued intended semantics used to locate a bug in an
instance of a single clause or a call with a delay annotation.

\bibliography{all}
%
%
%
\end{document}